\documentclass[prc,aps,nofootinbib,showpacs]{revtex4}
\usepackage{epsfig}
\begin{document}
\title{Nuclear level statistics: extending the shell model theory to
higher temperatures}
\author{Y. Alhassid$^{1}$, G.F. Bertsch$^{2}$, and L. Fang$^{1}$
}
\affiliation{$^{1}$Center for Theoretical Physics, Sloane Physics
Laboratory, Yale University, New Haven, CT 06520\\
$^{2}$Department of Physics and Institute of Nuclear Theory,
Box 351560\\
University of Washington
Seattle, WA 98915}

\date{March 18, 2003}

\def\be{\begin{equation}}
\def\ee{\end{equation}}
\def\Fe{$^{56}$Fe}
\def\Cr{$^{66}$Cr}
\def\Tr{{\rm Tr}}

\begin{abstract}

The Shell Model Monte Carlo (SMMC) approach has been applied to
calculate  level densities and partition functions to temperatures up to 
$\sim 
1.5  - 2$ MeV, with the maximal temperature limited by the size of the
configuration  space.  Here we develop an extension of the theory that
can be used to higher temperatures, taking into account the large
configuration space that is needed.  We first examine the
configuration space limitation using an
independent-particle model that includes both bound states
and  the continuum.  The larger configuration space is then combined with
the SMMC under the assumption that the effects on the partition function
are factorizable. The method is demonstrated for nuclei in the iron region, 
extending  the calculated partition functions and level densities up to $T 
\sim  4$ MeV. We find that the back-shifted Bethe formula has a much larger 
range  of validity than was  suspected from previous theory.  The present 
theory  also shows more clearly  the effects of the pairing phase transition 
on  the heat capacity.
\end{abstract}

\maketitle

\section{Introduction}

Consistent calculations of nuclear partition functions and level densities
are  required in nuclear astrophysics. The nuclear partition functions at a
finite  temperature are needed to understand pre-supernova  collapse
\cite{bethe79}  and for the calculation of stellar thermal
reaction  rates \cite{rauscher00}. Similarly, level densities determine the
statistical  neutron and proton capture cross-sections in nucleosynthesis
\cite{bbfh57}.

   The Fermi gas formula (also known as the Bethe formula) \cite{bethe36} for
the  nuclear level density can be derived from the partition function of
non-interacting  nucleons \cite{bm69}. Correlation effects, e.g., shell,
pairing  and deformation effects, are usually accounted for through empirical
modifications  of this formula.   In particular, the back-shifted Bethe 
formula  \cite{gc65,hm72} is found to describe well the experimental data if 
its  parameters, the single-particle level density parameter $a$ and 
back-shift 
parameter  $\Delta$, are fitted for each nucleus \cite{dilg73,il92}. However,
these  phenomenological approaches cannot be reliably used in nuclei for 
which 
experimental  data are not available, so there is a need for microscopic
theories that include interaction effects.  In mean field theory, pairing
can be treated in the BCS approximation, as was done in Ref.~\cite{go96}.
However, shape fluctuations are likely to be significant and they require
a theory beyond the mean field, e.g., the static path approximation 
\cite{la88}.   A theory that includes quadrupole fluctuations
in the static path approximation was proposed in Ref.~\cite{ag98}.  Like
the mean field theory, the static path approximation allows calculations
in large spaces. This was exploited by the authors of Ref.~\cite{ag98}
who presented results for large single-particle spaces including an 
approximate  treatment of unbound states. However, as implemented there, 
the theory includes neither pairing correlations, which are certainly of 
similar  importance, nor multipoles of the interaction beyond quadrupole.

A more systematic approach to calculate correlation effects is to use the
interacting shell model. In this approach, we define a many-particle 
configuration  space and treat in full the effective interaction within that 
space.  Here the shell model Monte Carlo (SMMC) method has rather favorable 
computational  properties, scaling  with
the number $N$ of single-particle orbitals as $N^4$, compared to
direct diagonalization methods, which scale exponentially.\footnote{Another 
method  in the literature is the use of moments of the
Hamiltonian to expand the level density in a sum of partitioned  Gaussians 
\cite{mf75,sg88,km96}.  For a similar approach using binomials instead of 
Gaussians  see Ref.~\cite{johnson01}.}
Even so, most applications of the SMMC have so far been limited to a single 
major  shell.  For medium mass nuclei, this restricts the reliability of the 
calculated  partition functions (and associated level densities) to 
temperatures  below $\sim  1.5 - 2$ MeV \cite{na97,al99,al00}. 

In this work, we will first explore the model space truncations in an
independent-particle approximation, extending the single-particle space to
include both the bound states and the continuum.  This
was studied within a semiclassical approximation in Ref.~\cite{tu79}. We
will develop the fully quantum-mechanical theory that includes the 
continuum\footnote{In  Ref.~\cite{ag98} the continuum was treated 
approximately  by enclosing the nucleus in a sphere whose radius $R$ is 
slightly  larger than the size of the nucleus. Our treatment is exact and 
corresponds  to the limit $R \to \infty$.}  in  Section~\ref{sec:independent}. 
  We find that in tightly bound nuclei, the continuum contribution is 
negligible  even at high temperatures (e.g., $T \sim 4$ MeV for $^{56}$Fe), 
while  in weakly bound nuclei the continuum contribution can become 
significant  already at low temperatures. In either case the continuum 
contribution  can be well approximated by considering just the contribution of 
the  narrow resonances, treated as broadened bound states.

The remaining task is to include interaction effects, as discussed in Section
\ref{sec:interaction}.  It is necessary to distinguish between the long-range 
part  of  the interaction, acting in the valence orbitals, and the short-range 
part,  involving  highly excited configurations. The short-range part is 
beyond  the scope  of the theory, and must be taken into account in defining 
an  effective long-range  interaction, but the latter can be accurately 
treated  within the SMMC  approach.
At higher temperatures, long-range interaction effects are weak and
corrections  to the entropy from the larger model space can be included in a
single-particle  approach.   We shall see that at high temperatures, the
interaction  effects on the partition function scale as $1/T$,
and the two temperature regimes can be joined smoothly. We find that the
extended  partition function and its associated level density are well
described  by the back-shifted Bethe formula up to $T \sim 4$ MeV.  This
value represents an upper limit for our underlying assumption that the
single-particle potential is fixed and independent of temperature.

The basic object we study is the canonical partition function defined by
\begin{equation}\label{partition}
Z(\beta) = \Tr e^{-\beta H}\;,
\end{equation}
where $\beta$ is the inverse temperature, $\beta= 1/T$, and the trace is
taken  at a fixed particle number (i.e., fixed number of protons and
neutrons).   We shall
find it convenient to measure the energy with respect to the ground
energy $E_0$, defining an excitation partition function $Z'$ as
\begin{equation}\label{excitation-partition}
Z' \equiv Ze^{\beta E_0} \;.
\end{equation}
Thus at zero temperature, the excitation partition function is equal to the
degeneracy  of the ground state $N_{\rm gs}$,
$Z'(T=0) = N_{\rm gs}$.

 The conversion from the partition function to the level density is discussed 
in  Section \ref{sec:level-density}.   While the results of the theory can be 
given  as numerical tables of $Z(T)$ and the corresponding level density 
$\rho(E)$,  it is also useful
to express the results in terms of the parameters of the back-shifted Bethe
formula, which provides a good description at not too low temperatures.  The 
inclusion  of a back shift requires
a particular parameterization of $Z'$, as discussed in Section 
\ref{sec:interaction}. 

\section{Orbital truncation effects in the independent-particle
approximation}\label{sec:independent}

  When the temperature is larger than about a third of the nucleon
separation energy, the unbound nucleon states  can no longer
be ignored.  To develop a theory that includes the continuum, we first
need to define precisely how we separate the contributions of the free
nucleons  and the nucleus to the partition function.  If the system is 
enclosed  in a box, the free nucleon
partition function is proportional to the volume of this box. The
contribution of the continuum to the nuclear partition function is found by
the  explicit subtraction of the free-particle contribution, and is
independent of the volume of the box, if it is large compared to the size of 
the  nucleus.  The free nucleon partition function we use here is the one 
obtained  when the nuclei are ignored entirely.\footnote{We also ignore the 
Coulomb  effects on the partition function, which are small but which 
immediately  introduce the complications of plasma theory.} 

In the independent-particle approximation, the single-particle
spectrum is calculated in an external potential well, representing
the nuclear mean-field potential. Assuming the well to be spherically
symmetric, the bound state energies $\epsilon_{nlj}$ depend only on the 
principal  quantum number $n$, orbital angular  momentum $l$, and total spin 
$j$.  For the positive energy continuum,
we  need the scattering phase shifts $\delta_{lj}(\epsilon)$ where $\epsilon$
is  the (positive) energy of the continuum states.  The change $\delta \rho$
in  the single-particle continuum level density in the presence of  the
potential  $V$ is found by subtracting the free particle
level density $\rho_0$ from the total density $\rho$:
\be
\delta\rho(\epsilon) = \rho(\epsilon)-\rho_0(\epsilon)=\sum_{l j}
(2j+1){1\over  \pi}
{d\delta_{lj}\over d\epsilon}
\;.
\ee

We denote the many-body grand canonical
partition function in the independent-particle approximation by $Z^{\rm
GC}_{\rm  sp}$, where  the subscript stands for ``single-particle.''  It is
given  by
\be\label{GC-partition}
\ln Z^{GC}_{sp} (\beta,\mu) = \sum_{nlj} (2j+1) \ln [1 +
e^{-\beta(\epsilon_{nlj}-\mu)}]  +
\int_0^\infty d\epsilon \delta\rho(\epsilon)\ln[1 + e^{-\beta(\epsilon-\mu)}]
\;,
\ee
where $\mu$ is the chemical potential.\footnote{In practice, 
Eq.~(\ref{GC-partition})   includes two separate contributions from protons 
and 
neutrons  with different chemical potentials $\mu_p$ and $\mu_n$.}
A more convenient form, avoiding the phase-shift derivative, may be
obtained by integrating the second term in the above equation by parts.
The  contribution from the integral endpoint at $\epsilon=0$ is evaluated 
using  Levinson's theorem, $\delta_{lj}(\epsilon=0)= n_{lj} \pi$,
where $n_{lj}$ is the number of bound $lj$ states.
The result is
\be
\int_0^\infty d\epsilon {d\delta_{lj}\over d\epsilon}
\ln[1 + e^{-\beta(\epsilon-\mu)}] = -n_{lj}\pi \ln (1+e^{\beta\mu}) +
\beta \int_0^\infty d\epsilon\; \delta_{lj}(\epsilon)f(\epsilon)
\;,
\ee
where $f(\epsilon)=[1 + e^{\beta(\epsilon-\mu)}]^{-1}$ is the Fermi-Dirac
occupation. The formula for the partition function can then be rewritten
\be
\ln Z^{\rm GC}_{sp}(\beta,\mu) = \sum_{lj} (2j+1)\left\{ \sum_{n}\ln \left[
{1  +
e^{-\beta(\epsilon_{nlj}-\mu)} \over 1 + e^{\beta \mu}}\right] +
{\beta \over \pi} \int_0^\infty d\epsilon\; \delta_{lj}(\epsilon) f(\epsilon)
\right\}
\;.
\label{full-sp}
\ee
This form is easier to use computationally because it avoids
numerical derivatives.

To calculate the partition function of a specific nucleus, we need to 
transform  the grand canonical partition to the canonical partition
function at fixed proton and neutron numbers.  In the saddle point 
approximation,  the corresponding correction is estimated from
the particle-number fluctuations in the grand-canonical ensemble, giving for 
the  canonical $Z(\beta)$
\begin{eqnarray}
Z (\beta)\approx \left(2 \pi \langle(\Delta N)^2\rangle \right)^{-1/2}
Z^{\rm GC} \; e^{-\beta\mu N}
\;,
\end{eqnarray}
or
\begin{eqnarray}\label{partition-N}
\ln Z' \approx \ln Z^{\rm GC} +\beta E_0 -\beta\mu N -{1\over 2}\ln \left( 2 
\pi   \langle(\Delta N)^2\rangle \right)
\;.
\end{eqnarray}
Here $\langle(\Delta N)^2\rangle$ is the variance of
the neutron or proton number and $Z' = Z e^{\beta E_0}$.  In 
Eq.~(\ref{partition-N}),   $\mu$ is calculated from $\langle \hat N \rangle =
\partial\ln  Z^{\rm GC}/
\partial \alpha= N$ (where $\langle \hat N \rangle$ is the average particle
number  in the grand-canonical ensemble and $\alpha=\beta\mu$), and the
variance  is found from
$\langle(\Delta N)^2\rangle={\partial^2\ln Z^{\rm GC}/\partial \alpha^2}$.
In  the independent-particle approximation these quantities are
\begin{eqnarray}\label{particle-number}
N = {\partial \ln Z^{\rm GC}_{\rm sp} \over \partial \alpha} = \sum_{lj}
(2j+1)\left[  \sum_{n} f_{nlj} +
{1 \over \pi} \int_0^\infty d\epsilon {d\delta_{lj}\over d\epsilon}
f(\epsilon)\right]
\;,
\end{eqnarray}

\be
\label{variance}
  \langle(\Delta N)^2\rangle = {\partial^2\ln Z^{\rm GC}_{\rm sp}\over
\partial \alpha^2} = \sum_{lj} (2j+1)
\left[ \sum_{n} f_{nlj}(1-f_{nlj}) +
{1 \over \pi} \int_0^\infty d\epsilon {d\delta_{lj}\over d\epsilon}
f (1-f)\right]
\;.
\ee
To avoid numerical derivatives, we can again use integration by parts and
rewrite  Eqs.
(\ref{particle-number}) and (\ref{variance}) as
  \begin{eqnarray}\label{particle-number1}
N = \sum_{lj} (2j+1)\left[ \sum_{n}[ f_{nlj}-f(0)] +
{\beta \over \pi} \int_0^\infty d\epsilon \; \delta_{lj}(\epsilon)f(1-f)
\right]
\;,
\end{eqnarray}
and
\begin{eqnarray}\label{variance1}
  \langle(\Delta N)^2\rangle  =  \sum_{lj} (2j+1)
\left\{ \sum_{n} \left[f_{nlj}(1-f_{nlj})  -  f(0)(1-f(0))\right]  \right.
\nonumber  \\ \left.  +
{\beta \over \pi} \int_0^\infty d\epsilon \;\delta_{lj}(\epsilon)f(1-f)(1-2f)
\right\}
\;.
\end{eqnarray}

\begin{figure}[t!]
\centerline{\epsfig{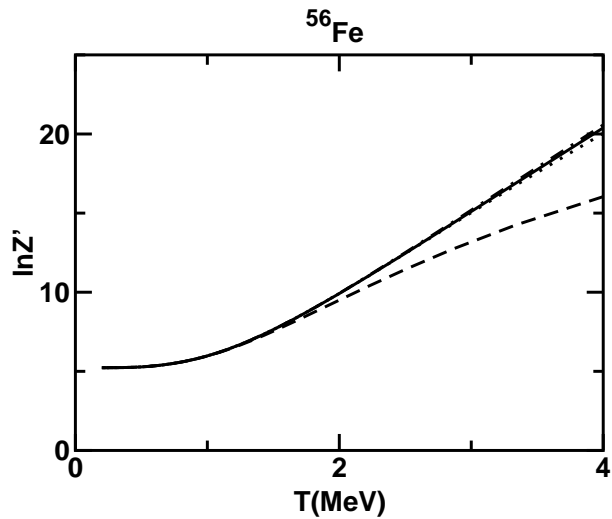}}
\caption{\label{fig1}
The logarithm of the canonical excitation partition function $Z'_{sp}$
(where the energy is measured with respect to the ground-state energy)
calculated for $^{56}$Fe is shown for the following single-particle spaces:
all bound states plus continuum as described by
Eqs.~(\ref{full-sp}), (\ref{partition-N}),
and (\protect\ref{variance1}) (solid line);
bound states and the narrow resonances listed in Table I (dashed-dotted
line);  and bound states only (dotted line).
The corresponding calculation with the space truncated to the $fpg$ major
shell is shown  as the dashed line.  The intercept at $T=0$ is $\ln N_{\rm 
gs}$, 
where $N_{\rm gs}=168$ is the number of states in the 
$(f^p_{7/2})^{-2}(p^n_{3/2})^2$ 
shell configuration.
}
\end{figure}

  To illustrate the method, we apply it to nuclei in the iron region. For the 
mean-field  potential we use a Woods-Saxon well $V(r)$ plus a spin-orbit
interaction
\begin{eqnarray} \label{Woods-Saxon}
V(r) + \lambda_{ls} (\ell \cdot
{\bf s})r_0^2 {1 \over r} {dV\over dr} \;,
\end{eqnarray}
where $V(r) =V_0 \left[ 1 +\exp(r-R_0)/a\right]^{-1}$ with
$R_0=r_0A^{1/3}$. We choose the parameterization of  Ref.~\cite{bm69},
where $V_0= -51 + 33 {N-Z \over A}$ MeV, $a=0.67$ fm, $r_0=1.32$ fm, and
$\lambda_{\ell  s}=- 0.44$. We have calculated the single-particle spectrum 
and  scattering phase shifts in this potential well, and used them to compute 
the  canonical partition function from Eqs.~(\ref{partition-N}), 
(\ref{full-sp})  and (\ref{variance1}).  The solid line in 
Fig.~\ref{fig1} shows the
logarithm  of the canonical excitation partition function $\ln Z_{\rm sp}'$
for  the nucleus $^{56}$Fe when the full single-particle space (bound states 
plus  continuum) is taken into account.

\begin{table}[h!]
\caption{\label{TableI} Neutron orbital energies and resonance energies (in
MeV)  in \Fe, calculated
with the Woods-Saxon potential of Ref.~\protect\cite{bm69}.}
\begin{tabular} {cc|cc}
Orbital type & $l,j$  & $\epsilon$ (\Fe) & $\epsilon$ (\Cr) \\
\tableline
bound & $f_{7/2}$ &  -12.68 & -10.2\\
& $p_{3/2}$ & -8.65 & -6.7\\
& $p_{1/2}$ & -6.56 & -5.1\\
& $f_{5/2}$ & -6.34 & -5.2\\
& $g_{9/2}$ & -3.13 & -1.7\\
& $d_{5/2}$ & -0.41 & \\
& $s_{1/2}$ & -0.26 & \\
\tableline
resonance & $d_{5/2}$ &  & 0.35 \\
     & $g_{7/2}$ & 4.95 & 4.67 \\
     & $h_{11/2}$ & 6.22 & 6.62 \\
\end{tabular}
\end{table}

We next investigate various truncations within the independent-particle 
model. 
Of  particular interest is the truncation to the same model space that is 
used 
in  the SMMC approach. Most of the SMMC calculations so far have been
restricted  to the bound states
in a single major shell.  For example calculations in the iron region were
restricted  to the complete $fpg_{9/2}$ major shell, which includes the 
active  orbitals
$f_{7/2},p_{3/2},p_{1/2},f_{5/2}$ and $g_{9/2}$.  The orbital energies
[determined  in a Woods-Saxon well with spin-orbit interaction 
(\ref{Woods-Saxon})] 
are given in Table I.

The partition function calculated in this truncated
space (in the independent-particle approximation) is compared with the full
space  calculation [Eqs. (\ref{full-sp}), (\ref{partition-N}) and
(\ref{variance1})]  in Fig.~\ref{fig1}.
One sees that the truncation to a single major shell becomes
problematic at temperatures above $\sim 1.5$ MeV.

To assess the importance of the continuum, we also show in
Fig.~\ref{fig1} the result of the
calculation keeping all bound states but neglecting the continuum integral.
One sees that the continuum contribution is not very
significant  for a tightly bound nucleus such as \Fe.

\begin{figure}[t]
\centerline{\epsfig{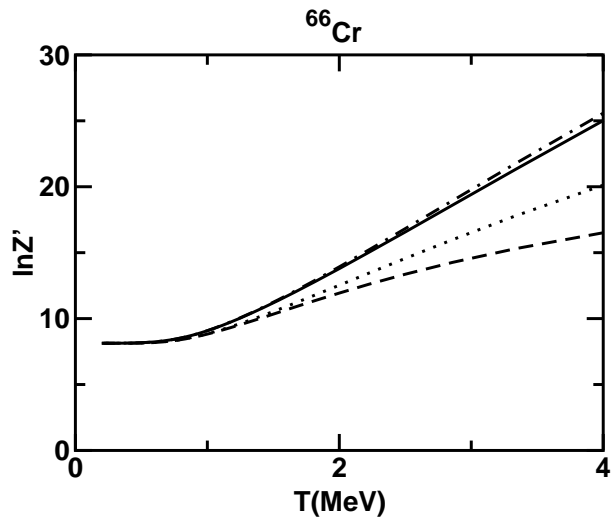}}
\caption{\label{fig2}
Similar to Fig.~\protect\ref{fig1} but for the nucleus \Cr.
}

\end{figure}

However this situation changes in
a nucleus with a small neutron separation energy.
In Fig.~\ref{fig2} we compare the calculations in the full single-particle
space  and in the truncated space that includes all bound states for the
nucleus  \Cr.  Here the separation
energy of the highest occupied orbital is 1.7 MeV.  One sees that the
continuum effects become significant  for $T \agt 1$ MeV.

Rather than ignoring the continuum completely, one could also consider
the approximation of including only narrow resonances, treating them
as broadened discrete states.  Indeed, the derivative of the scattering phase 
shift  $\delta_{lj}$ in  the vicinity of a narrow resonance follows a 
Lorentzian 
\begin{equation}
{d\delta_{lj}\over d\epsilon} \approx {\Gamma^r_{lj}/2 \over
(\epsilon-\epsilon^r_{lj})^2 +(\Gamma^r_{lj}/2)^2}
\;,
\end{equation}
where $\epsilon^r_{lj}$ and $\Gamma^r_{lj}$ are the centroid and width of a
resonance  with quantum numbers $lj$.
If the resonance width $\Gamma^r_{lj}$ is much smaller than $T$,
the integral in (\ref{GC-partition}) can be evaluated to give a contribution 
to  $\ln  Z_{sp}^{GC}$ as though the resonance were a discrete state with 
energy  $\epsilon^r_{lj}$.

We identify the resonances from the energy dependence of the phase shifts 
$\delta_{lj}$  in different channels $lj$, and take their energies where the 
phase  shifts  ascend through $\pi/2$. The  channels with resonances and their 
energies  are displayed in Table I.  The partition function calculated using 
bound  states and narrow resonances  is
shown  as the dotted line in Fig.~\ref{fig2}.
We see that the resonance approximation is good at least to
$T\sim 3.5$ MeV.

We now summarize the validity of the various approximations, requiring
that the theory be accurate up to temperatures  of $T\sim 4$ MeV.
For nuclei along the stability line, it is only necessary to include
bound states.  If one goes to nuclei that have much lower separation
energies, the continuum contribution can be significant but the resonance
approximation is likely to be adequate. It should be emphasized throughout 
this  section we have assumed a fixed mean-field potential. Above $T \sim 4$ 
MeV,  that approximation  becomes unreliable.

\section{Interaction effects}\label{sec:interaction}

  We next discuss interaction effects on the partition function.  The
ground-state binding energy is of course very sensitive to the interaction,
but  this will not be immediately visible in the excitation partition
function. Let us denote by $Z'_{v,tr}$ the partition function calculated with
interactions  but in a truncated single-particle space.  We
argue that the correction due to a larger model space will be largely
additive  in the logarithm of the partition function, so that the way to 
combine  a small-space calculation of interaction effects with a large-space 
calculation  of the independent-particle partition $Z'_{sp}$ is with the 
formula 
\be\label{extended-partition}
\ln Z'_v = \ln Z'_{v,tr} +\ln Z'_{sp} -\ln Z'_{sp,tr} \;,
\ee
where $Z'_{sp,tr}$ is the excitation partition function (in the absence of 
interactions)  in the same truncated
single-particle space in which $Z'_{v,tr}$ is calculated.

Eq.~(\ref{extended-partition}) cannot be justified rigorously, but we can 
motivate  it in several ways.  It is certainly true that the interaction 
corrections  approach zero at high temperature.  From finite-temperature 
many-body  perturbation theory, it is seen that the leading corrections to the
independent-particle Hamiltonian $H_0$ are additive in the logarithm of
the partition function, as is assumed in Eq.~(\ref{extended-partition}). The 
explicit  formula is
\be\label{perturbation}
\ln \Tr\ \exp(-\beta K)-\ln \Tr\ \exp(-\beta K_0) =
- \int_0^\beta d\tau\langle V_I(\tau)\rangle
+ \frac{1}{2}\int_0^\beta d\tau_1 \int_0^\beta d\tau_2 \langle T_\tau
V_I(\tau_1)V_I(\tau_2)
\rangle_{\rm conn} +\ldots \;,
\ee
where $K=H -\mu \hat N$, $K_0=H_0 -\mu \hat N$, $V_I$ is the two-body 
interaction  in the interaction picture (with respect to $K_0$), $T_\tau$ 
denotes  time ordering and $\langle \ldots\rangle$ denotes a thermal average 
with  respect to $K_0$.  Each integration in (\ref{perturbation}) gives a 
factor  $1/T$.
  In cases where analytic expressions can be derived (e.g., hard core gas and 
Coulomb  gas), it may be verified that the interaction correction falls off as 
$T^{-1}$  or more strongly at
high $T$.  At low temperatures, the extended space is not thermally
occupied and only plays an indirect role, renormalizing the
interactions in the smaller space.   Since the SMMC is applied with
renormalized interactions that are appropriate for the smaller space, 
there is no additional correction when the larger space is considered
explicitly.  With the correct limiting behavior at both
low and high temperatures, we have a better theory for the complete
temperature dependence.

We now give a brief summary of the SMMC calculations to be used for
$Z'_{v,tr}$ in Eq.~(\ref{extended-partition}).  The interaction is taken to 
be  the same as in Ref.\ \cite{na97}. It is separable and surface-peaked,
acting  between orbitals of a major shell (here the $fpg_{9/2}$ orbitals). 
The  strength of the surface-peaked interaction  is determined 
self-consistently  and renormalized appropriately
for each multipole.  In addition, there is a monopole pairing interaction,
whose  strength is determined from odd-even mass differences.  The
single-particle Hamiltonian is the same as in Section \ref{sec:independent}.
The outputs of the  SMMC computation are canonical expectation values of 
various  observables. For our purposes here, the most important
quantity is the canonical thermal energy $E(\beta)$, calculated as the
expectation of the Hamiltonian with number-projected SMMC densities.
The partition function is then found by integration
\be\label{canonical-Z}
\ln Z(\beta) = \ln Z(0) - \int_{0}^\beta d \beta' E(\beta') \;.
\ee
Here $Z(0)$ is the partition function at $T\to \infty$ and is equal to
the total number of many-particle states in the model space. In addition,
we need to determine the ground-state energy $E_0$ in order to find the
excitation partition function, Eq.~(\ref{excitation-partition}).
This is done by extrapolating the calculated $E(\beta)$ to large
$\beta$.

\begin{figure}[t]
\centerline{\epsfig{figure=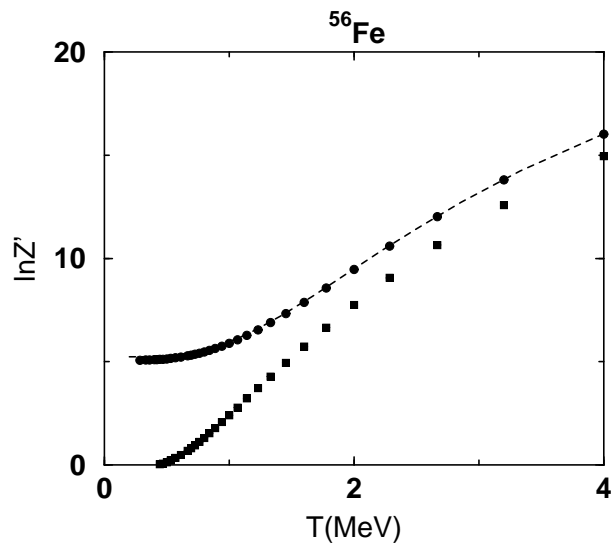,width=8 cm,clip=}}
\caption{\label{fig3}
 The excitation partition function of $^{56}$Fe using the SMMC method in the 
truncated  space ($fpg_{9/2}$). The partition function $Z'_{v,tr}$ in the 
presence  of the residual interaction (squares) is compared to the SMMC 
partition  function $Z'_{sp,tr}$ in the absence of interactions (circles).  
The  dashed line is the partition function of the independent-particle model 
calculated  in the saddle point approximation using Eqs.~(\ref{partition-N}), 
(\ref{full-sp})   and (\ref{variance1}).
}
\end{figure}

The SMMC results for the nucleus $^{56}$Fe are shown in Fig.~\ref{fig3}, with 
$\ln  Z'_{v,tr}$ plotted as a function of $T$ (squares).  The
curve  starts out flat,
changing to a linear increase over some range of $T$, and reaching a plateau
beyond  the highest temperature plotted.  The initial flat behavior is
associated with the gap in the energy spectrum between the ground state and 
the  first excited state,  while the linear regime corresponds to the Bethe 
formula.    The saturation  is due to the truncation of the space, and its 
onset  marks the limit of validity of the truncated SMMC calculation.   We 
next  repeat the SMMC calculation with the two-body interaction turned off. 
The  results are shown by the circles in Fig.~\ref{fig3}.
Notice that the difference between the two partition functions becomes
small as the temperature increases, confirming the analysis in
an earlier part of this section [see in particular Eq.~(\ref{perturbation})].

Before calculating the partition function in the extended model space,
we also compare in Fig.~\ref{fig3} two ways of calculating the 
non-interacting  partition function, from Eq.~(\ref{partition-N}) (dashed 
line)  or from the SMMC with the interaction turned off (squares). In the 
latter  method, particle-number projection is carried out exactly,
while in Eq.~(\ref{partition-N}) this is done in the saddle point
approximation.   One can see that the differences are insignificant for our 
purposes. 

\begin{figure}[t]
\centerline{\epsfig{figure=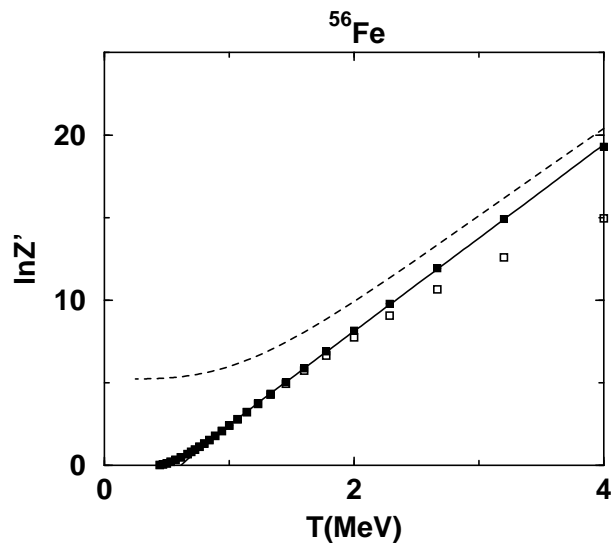,width=8 cm,clip=}}
\caption{\label{fig4}
Extended partition function for $^{56}$Fe, calculated from
Eq.~(\protect\ref{extended-partition})  shown
as solid squares.  The solid line is a fit to
Eq.~(\protect\ref{BBF-partition})  with $a=5.87$ MeV$^{-1}$ and $\Delta=1.03$
MeV.  Shown for comparison are the SMMC
(open squares) and the full space single-particle (dashed line) results.
}
\end{figure}

We now combine the various terms in Eq.~(\ref{extended-partition}) to get our
extended-range  partition function.  The result is
shown in Fig.~\ref{fig4} by the solid squares.
For comparison, the SMMC result (open squares) and the full space 
independent-particle  model result (dashed line) 
are shown as well.  Remarkably, the extended $\ln Z'$ is seen to be close to
a linear function of $T$ over most of the range, unlike either of its
constituents.  In the saddle point approximation, the dominating term in $\ln 
Z'$  is linear in $T$ with whose slope is the parameter $a$ in the level 
density  formula.  However, to compare with the back-shifted Bethe formula 
[see  Eq. (\ref{BBF}) in Section \ref{sec:level-density}], more care is needed 
in  treating the sublinear terms.  Starting from the Darwin-Fowler expression
for the partition function of non-interacting fermions [see Eq.~(2B-9) in 
Ref.~\cite{bm69}], 
and including the proton and neutron number fluctuation terms from 
Eq.~(\ref{partition-N}), 
one can derive the following expression for $Z'$,
\begin{equation}\label{Bethe-partition}
\ln Z' \approx a T -\ln(6 a T/\pi) \;,
\end{equation}
where $a=\pi^2 g/3$ and g is either the proton or neutron single-particle
level density at the Fermi energy (assumed to be equal).
Within the saddle-point approximation, Eq.~(\ref{Bethe-partition}) is
equivalent to the simple Bethe formula
for  the level density  (see Section
\ref{sec:level-density}).  As mentioned in the introduction, a better 
phenomenological  
description of the measured level density
is often obtained by shifting the ground state energy by an
amount  $\Delta$ in the Bethe formula. We show in Appendix A that the
corresponding modification of the partition function is to introduce
a third term in Eq.~(\ref{Bethe-partition}),
\begin{equation}\label{BBF-partition}
\ln Z' \approx a T -\ln(6 a T/\pi) -\Delta/T\;.
\end{equation}
The solid line in Fig.~\ref{fig4}
is a fit of (\ref{BBF-partition}) to $\ln Z'$ for $1 \alt T \alt 4$ MeV
with  $a=5.87$ MeV$^{-1}$ (corresponding to the expression
$a= A/K$ with $K=9.5$ MeV) and $\Delta=1.03$ MeV.  The fit does remarkably 
well  in describing the
$T$-dependence of $\ln Z'$ for temperatures between $T \sim 1$ MeV and $T 
\sim  4$ MeV (the rms of the deviation from the calculated $\ln Z'$ is 
$0.05$).  The functional form
(\ref{BBF-partition})  of $\ln Z'$ is the essence of the (back-shifted) Bethe
level  density formula; the result observed here implies that this formula is
useful  to considerably higher temperatures than was apparent in earlier
calculations.  At high temperatures, the dominant term in
(\ref{BBF-partition})  is linear in $T$. However, the other two terms are
necessary  to obtain a value of $a$ that is comparable to the value extracted
directly  from the level density, which will be the subject of Section 
\ref{sec:level-density}.  

 In the independent-particle model, $\ln Z'$ is approximately linear in $T$ 
only  for $T \agt 2$ MeV with a slope of $a\approx 5.25$ MeV$^{-1}$, i.e., 
$K\approx  10.7$. This value of $K$ is smaller than the Fermi gas 
value~\cite{bm69}  $K\approx 15$ but still larger than the value we found 
above  when correlations are taken into account. 

\begin{figure}[t!]
\centerline{\epsfig{figure=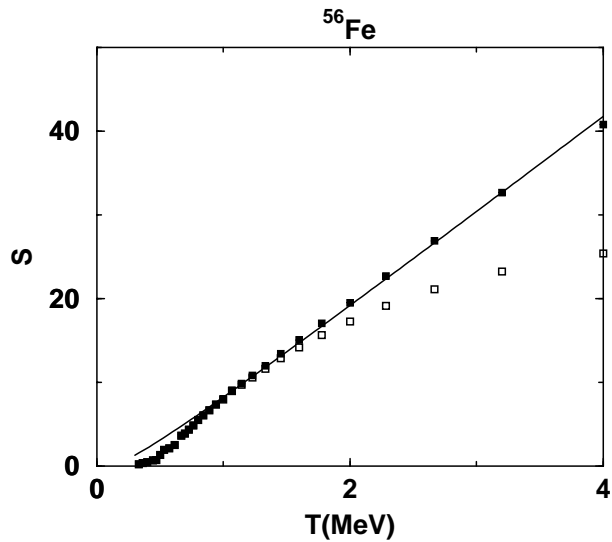,width=8 cm,clip=}}
\caption{\label{fig5}
The extended entropy of $^{56}$Fe (solid squares). The solid line is a fit to
Eq.~(\protect\ref{BBF-entropy})  with $a=5.82$ MeV$^{-1}$. The open squares
describe  the SMMC entropy.
}
\end{figure}

The canonical entropy can be calculated from $S=-\partial F/\partial T$ where
$F=  -T \ln Z'$ is the canonical free energy.
Using the empirical formula (\ref{BBF-partition}), this entropy is given by
\be\label{BBF-entropy}
S \approx 2 a T -\ln(6 a T/\pi) -1 \;,
\ee
and is independent of the back-shift parameter $\Delta$. Thus the parameter
$a$  can be determined by a single parameter fit of the extended entropy to
(\ref{BBF-entropy}).  The leading order term in the entropy is linear
in  $T$ and the last two terms in (\ref{BBF-entropy}) originate in the
particle  number fluctuations.
The microscopic calculation of the extended entropy can
be  done without taking numerical derivate with respect to $T$, as is 
explained 
in  Section \ref{sec:level-density}. Fig.~\ref{fig5} shows a fit of the
empirical  formula (\ref{BBF-entropy}) (solid line) to the extended entropy 
in  the range $T \sim 1 - 4$ MeV
(symbols)  with $a=5.82$ MeV$^{-1}$ (the rms deviation is $\approx 0.3$).  
This  value of $a$ is in agreement with the
value  found from $\ln Z'$. The fit is good and only at $T
\sim  4$ MeV, do we start to observe a small deviation.
 
\begin{figure}[t!]
\centerline{\epsfig{figure=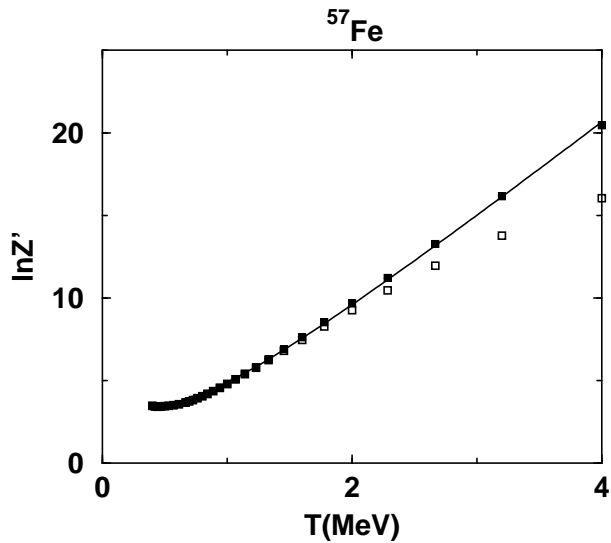,width=8 cm,clip=}}
\caption{\label{fig6}
The extended partition function (solid squares)  for the odd-even nucleus
$^{57}$Fe  is compared with the SMMC partition function (open squares). The 
solid  line is a fit to the back-shift formula (\ref{BBF-partition}) with 
$a=6.04$  MeV$^{-1}$ and $\Delta=-1.29$ MeV.
}
\end{figure}

  We have demonstrated our method for an even-even nucleus $^{56}$Fe, but
it should be equally applicable to odd-even and odd-odd nuclei. As an
example we show in Fig.~\ref{fig6}  the logarithm of the excitation partition 
function  for  $^{57}$Fe. The solid line is a fit to (\ref{BBF-partition}) 
with  $a=6.04$ MeV$^{-1}$  and $\Delta=-1.29$ MeV in the range $1 \alt T \alt 
4$  MeV (the rms deviation is $\approx 0.08$). A
qualitative  difference is that unlike the case of $^{56}$Fe, $\ln Z'$ does
not  approach zero at low temperatures. This is due to the spin degeneracy of
the  ground state of an even-odd nucleus and the much smaller energy gap to
the  first excited state.

\section{Level densities}\label{sec:level-density}

We now determine the level densities associated with the extended
partition function of the previous section.
The level densities are calculated from the partition function as in
Refs.~\cite{na97}  and \cite{al99} using the saddle-point formula
\begin{equation}\label{level-density}
\rho(E) = (2 \pi \beta^{-2} C)^{-1/2} e^S \;,
\end{equation}
where $S=\ln Z +\beta E= \ln Z' +\beta(E-E_0)$ is the canonical entropy and
$C = -\beta^{-2} d E/d\beta$ is the canonical heat capacity.  The thermal
energy  in (\ref{level-density}) is a function of $\beta$ determined by
\be\label{E-beta}
E - E_0 = -{\partial \ln Z' \over \partial \beta} \;.
\ee
 Using the expression
(\ref{extended-partition})  for the extended partition function and
Eq.~(\ref{E-beta}),  we have
\be\label{excitation-energy}
E - E_0 = (E_{v,tr} - E^0_{v,tr}) + [E_{sp} - (E_{sp,tr} + E_{\rm core}^0)]
\;,
\ee
where $E_{\rm core}^0 = E_{sp}^0 - E_{sp,tr}^0$ is the $T=0$ energy of the
core  (i.e., the shells below the valence shell). The notation used here for
the  various energies follows the same notation used for the partition
functions  in (\ref{extended-partition}) and the subscript or superscript
``0''  denotes the ground state in the corresponding model space.

  Similarly, we find for the extended entropy and heat capacity
\begin{eqnarray}\label{S-C}
S &= & S_{v,tr} + S_{sp}-S_{sp,tr} \nonumber \\
C & = & C_{v,tr}+C_{sp}-C_{sp,tr}
\;.
\end{eqnarray}

  Eq.~(\ref{excitation-energy}) is used to determine the relation between the
excitation  energy and $\beta$ in the calculation of the extended level
density.  When the SMMC approach is used to calculate interaction effects in
the  truncated space, $E_{v,tr}$ is calculated directly from the expectation
value  of the Hamiltonian and $E^0_{v,tr}$ is found by extrapolation to 
$\beta 
\to  \infty$. The partition $Z_{v,tr}$ (and therefore the entropy $S_{v,tr}$)
is  found by integration using (\ref{canonical-Z}), while the heat capacity
$C_{v,tr}$  is found by a numerical derivative of the energy.

It remains to determine the respective quantities in the
independent-particle  model (both in the full and truncated spaces). Here
there  is a technical complication in that the needed quantities are 
canonical.  For example,  we need to calculate the canonical energies $E_{sp}$ 
and  $E_{sp,tr}$.  In principle, one can use (\ref{E-beta}) where $Z$ is 
calculated  in  the independent-particle model. However, since $Z$ is a 
canonical  partition  function, the partial derivative with respect to $\beta$ 
must  be evaluated  at constant particle number $N$ rather than at constant 
$\alpha$  ($\alpha=\beta  \mu$). The explicit expressions for $E_{\rm sp}$ and 
$C_{\rm  sp}$ in terms of logarithmic derivatives of the grand-canonical 
partition  function are derived in Appendix B. In Appendix C we express these 
logarithmic  derivatives in terms of the corresponding single-particle 
spectrum  and scattering phase shifts.

\begin{figure}[t]
\centerline{\epsfig{figure=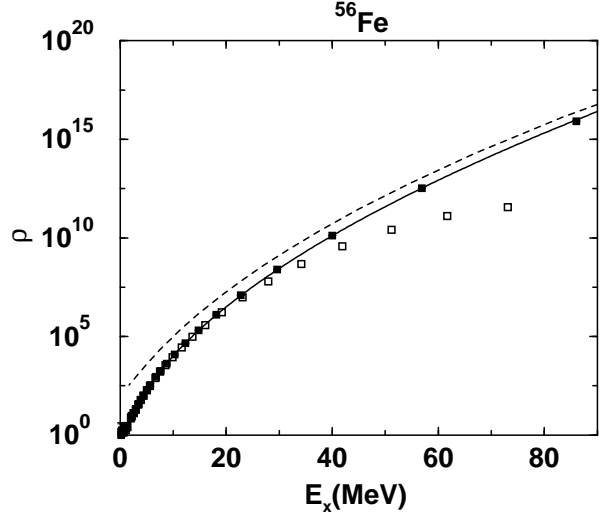,width=8 cm,clip=}}
\caption{\label{fig7}
Level density of $^{56}$Fe, calculated with the extended partition
function (solid squares). The solid line is a fit to the back-shifted Bethe
formula  (\protect\ref{BBF}) and describes well the extended level density in
the  full excitation energy range shown in the figure. The SMMC level density
in  the truncated $fpg_{9/2}$ shell is shown by open squares and the 
independent-particle   model level density is shown by the dashed line.
}
\end{figure}

  The extended level density of $^{56}$Fe is shown in Fig.~\ref{fig7} (solid
squares)  as a function of excitation energy, $E_x=E-E_0$. For comparison, 
the 
SMMC  level density calculated in the truncated space (i.e., $fpg_{9/2}$ 
shell)  is shown  by open squares, while the level density  calculated in the
independent-particle  model (no truncation) is shown by the dashed line.

To derive the back-shifted Bethe formula for the level density, we evaluate
the  inverse Laplace transform of $Z'$ in Eq.~(\ref{Bethe-partition}) in the
saddle  point approximation (separating out the particle number fluctuation
term  as a pre-exponential factor).  We find
\be\label{BBF-T}
\rho \approx (2 \pi)^{-1/2} \left(-{\partial E\over \partial
\beta}\right)^{-1/2}  \left({\pi \over 6 a T }\right) e^{ aT - \Delta/T +
\beta  (E-E_0)} = {\sqrt{\pi}\over 12} a^{-3/2} T^{-5/2} e^{2aT} \;,
\ee
where the temperature is determined by the excitation energy $E_x=E-E_0$
through
\begin{equation}\label{quadratic}
E_x \approx a T^2 +\Delta \;.
\end{equation}
Relation (\ref{quadratic}) generalizes the usual Bethe relation $E_x=a T^2$
to  include an empirical offset $\Delta$ to the ground-state energy. This
offset  originates in correlation effects.

When expressed in terms of the excitation energy, Eq.~(\ref{BBF-T}) is just
the  back-shifted Bethe formula
\be\label{BBF}
\rho(E_x) = {\sqrt{\pi}\over 12} a^{-1/4}(E_x-\Delta)^{-5/4}\exp
[ 2\sqrt{a(E_x-\Delta)}]  \;.
\ee
This formula has
two fit parameters, the single-particle level density parameter $a$ and the
back-shift  parameter $\Delta$. Fitting (\ref{BBF}) to the calculated 
extended 
level  density in the range $5 \alt E_x \alt 60$ MeV, we find $a= 5.9$ 
MeV$^{-1}$ 
and  $\Delta= 1.35$ MeV (with a $\chi^2$ per degree of freedom below 1), in 
overall  agreement with the values found in Section
\ref{sec:interaction}  by fitting (\ref{BBF-partition}) to $\ln Z'$.  These
values  of $a$ and $\Delta$ are also in close agreement with the earlier
values  found by fitting to the SMMC level density in the energy range  $5 
\alt  E_x  \alt 16$ MeV, $a= 5.85$ MeV$^{-1}$ and $\Delta= 1.37$ MeV. However, 
now  the back-shifted  Bethe formula (\ref{BBF}) describes the extended level 
density  up to  much higher values of $E_x$ ($\sim 90$ MeV).\footnote{A 
temperature  of T=4 MeV for $^{56}$Fe corresponds to $E_x \approx 86$ MeV.}   
Thus  the validity of this formula is now
extended  to significantly higher excitation energies than previously 
thought. 
Furthermore,  the specific values of the parameters $a$ and $\Delta$ found in
previous  SMMC calculations~\cite{al99} are approximately valid in this
extended  energy range.

\begin{figure}[t]
\centerline{\epsfig{figure=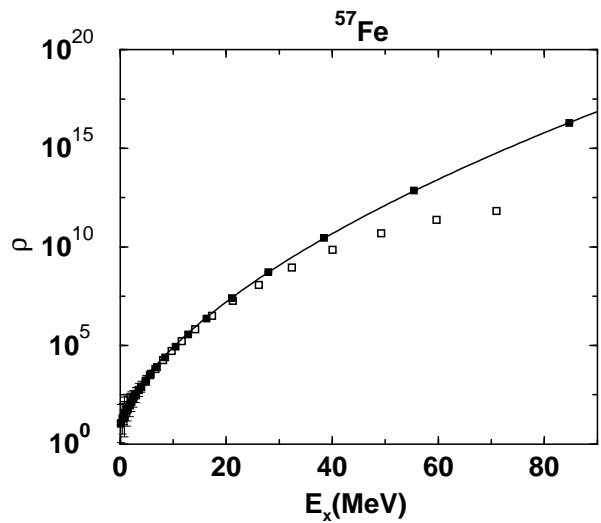,width=8 cm,clip=}}
\caption{\label{fig8}
Extended level density (solid squares) in comparison with the SMMC level
density  (open squares) for $^{57}$Fe. The solid line is a fit to the
back-shifted  Bethe formula (\protect\ref{BBF}).
}
\end{figure}

Fig.~\ref{fig8} shows the level density of the odd-$A$ nucleus $^{57}$Fe,
where  we find in the extended range $a= 6.05$ MeV$^{-1}$ and $\Delta= -1.3$
MeV.  These values are comparable to the values determined from $\ln Z'$ in
Section  \ref{sec:interaction}.

The good fits we find to the back-shifted formulas
(\ref{BBF-partition})  and (\ref{BBF}), for the logarithm of the partition
function  and the level density, respectively,  imply that the excitation 
energy  can be approximated by the
quadratic  expression (\ref{quadratic}) for not too low temperatures. 

Experimentally, the statistical properties of nuclei at higher temperatures
have  been deduced from heavy ion reactions.  If the nuclei are equilibrated
when  they decay, there is a direct relation between the
kinetic energy distributions of the decay products and the temperature
of the daughter nucleus.  Experimental results have been reported
indicating that the quadratic dependence of $E_x$ on temperature needed to
be  modified at high temperature~\cite{go89}. However in other experiments 
only  a slight deviation was found~\cite{ca00}. These latter experiments 
concluded  a weak  decrease of $a$ with temperature and even a constant $a$ 
could  not be excluded,  in agreement with our results.  

\section{Heat capacity and the pairing phase transition}

The back-shifted parameterization works well for temperatures $T \agt 1$ MeV, 
but  not at lower temperatures.  The deviations can be clearly seen in the 
canonical  heat capacity as a function of the temperature.  There can be
a peak in this function that is associated with the pairing phase transition,
but the back-shifted parameterization permits only a monotonic function
rising linearly with temperature. 
Recently, experiments in rare-earth nuclei have been reported  in
which effects of pair breaking can be seen in the canonical heat 
capacity~\cite{me99,sc01}.    The authors measured the level density,
constructed a canonical partition function from their data, and used it
 to deduce the heat capacity as a function of temperature. 

Such low temperature characteristics, especially those requiring numerical
derivatives, are nontrivial to calculate in the SMMC, but in Ref.~\cite{li01} 
a method was found to accurately calculate the heat
capacity in the temperature range of interest. A slight enhancement over
the Fermi gas heat capacity was found in even iron nuclei around $T \sim 0.8$
MeV.  This enhancement is more pronounced for the neutron-rich iron isotopes
and  is correlated with the reduction in the number of $J=0$ neutron pairs
with  temperature. However, because the calculation was restricted to a 
finite 
space  ($fpg_{9/2}$ shell), the calculated heat capacity reached a maximum 
around  temperatures  of $\sim 1.5$ MeV and the effect of the pairing phase 
transition  was less pronounced.

\begin{figure}[t]
\centerline{\epsfig{figure=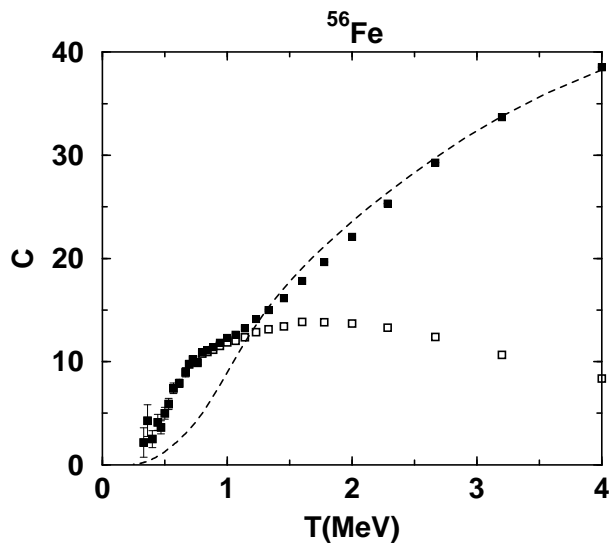,width=8 cm,clip=}}
\caption{\label{fig9}
The extended heat capacity of $^{56}$Fe versus temperature, showing a bump
around  the pairing transition temperature (solid squares). The SMMC heat
capacity  is shown by the open squares. The dashed line is the heat capacity
obtained  from the independent-particle model (bound states plus continuum).
}
\end{figure}

Here we use the extended theory to display the heat capacity in
a  much broader temperature range (i.e., up to $\sim 4$ MeV), where
deviations from the basic linear dependence become obvious. The effect of
the  pairing transition on the heat capacity is now clearly observed
even  in the iron isotopes with a smaller number of neutrons in which the 
effect  was  not noticeable before, e.g., $^{56}$Fe. Fig.~\ref{fig9} shows the
extended  heat capacity of  $^{56}$Fe calculated from (\ref{S-C}) and
(\ref{C-sp})  (solid squares) in comparison with the independent-particle 
heat 
capacity  calculated in the full space (solid line). Contrary to the SMMC 
heat 
capacity,  the extended heat capacity continued to increase monotonically 
with 
temperature  as does $C_{sp}$. However, it also exhibits a `shoulder' at low
temperatures  that is a clear signature of the pairing phase transition and 
is 
qualitatively  similar to the measured heat capacity in even-even rare-earth
nuclei~\cite{sc01}.

\section{Conclusion}

  In this work we have combined the partition function of the
 independent-particle model in the full space with the partition function of 
the  interacting shell model calculated in a small space by the SMMC method.  
This  enables us to extend the SMMC calculations of partition functions and 
level  densities up to significantly higher temperatures. The results
for iron nuclei suggest that the back-shifted Bethe formula  (\ref{BBF}) 
(with  temperature-independent parameters) is valid
over a wide range of temperatures extending from $T \sim 1$ MeV to $T \sim 4$ 
MeV. 
The two parameters of the formula, the single-particle level density 
parameter  $a$ and the back shift $\Delta$, agree well with the empirical 
values.  A corresponding empirical formula (\ref{BBF-partition}) that 
accommodates  correlation effects describes well the logarithm of the 
excitation  partition function.  

Expressed in the form  $a=A/K$, the theoretical value for $^{56}$Fe
is $K=9.5$ MeV.  This is considerably smaller than the Fermi-gas
value of $K \approx 15$ MeV. It is also smaller than the value $K \approx 
10.7$  MeV extracted from the high temperature slope of the logarithm of the 
independent-particle  partition function, indicating the importance of 
correlation  effects.  
In the literature, effective values of the parameter $a$ are often defined
from  relations that are valid in the Fermi gas limit (and that ignore 
corrections  due to particle-number fluctuations); e.g., $E_x=a T^2$,
$S=2aT$  and $S^2=4 a E_x^2$.  In the presence of shell and correlation 
effects,   these effective values of $a$ usually differ from each other and 
exhibit  a temperature dependence even at lower temperatures. 
Using the empirical relation (\ref{BBF-partition}) for the partition 
function,  we find a constant value of $a$ that is similar to the value 
extracted  from the level density itself. 

The effects of interactions and unbound states were also taken into 
account in Ref.~\cite{ag98}.  The authors parameterized their results
with a temperature-dependent $K$, and found it to vary significantly,
unlike our $K$.  However, their work only included part of the
interaction, omitting in particular the pairing interaction (which is largely
responsible for the back shift), and their treatment of the continuum was 
approximate.  The variation they observe at low temperatures $T \sim 1 - 2$ 
MeV  is at least partly due to definitions of $K$ that are based on the Fermi 
gas  limit.

For temperatures below $\sim 1$ MeV, the pairing effects are strong and 
beyond 
the range of applicability of simple parameterizations.  This 
can be seen in the canonical heat capacity if it is displayed
over a large temperature range, as we did using our extended theory.

\section*{Acknowledgments}

We thank G. Fuller and R. Vandenbosch for useful conversations. One of us
(Y.A.)  would like to acknowledge the hospitality of the Institute of Nuclear 
Theory  in  Seattle where part of this work was completed. This work was 
supported  in part  by the Department of Energy under Grants
DE-FG06-90ER40561 and DE-FG02-91ER40608.

\appendix
\section*{appendix A}

In this Appendix we derive the empirical back-shifted formula 
(\ref{BBF-partition})  for the excitation partition function. The partition 
function  is obtained from the level density by a Laplace
transform

\be\label{Laplace}
Z'(\beta) = \int_0^\infty \rho(E_x) e^{-\beta E_x} dE_x \;.
\ee
Substituting the back-shifted Bethe formula (\ref{BBF}) for $\rho(E_x)$ in
Eq.~(\ref{Laplace}),  and defining $x=E_x -\Delta$, we have
\be\label{BBF-integral}
Z'= {\sqrt{\pi}\over 12} a^{-1/4} e^{-\beta\Delta} \int^\infty x^{-5/4}
e^{2\sqrt{ax}  -\beta x}  \;.
\ee
We evaluate the integral in (\ref{BBF-integral}) in the saddle point
approximation.  The saddle point is given by $x_0=a T^2$, and we find
\be
Z' \approx  {\pi \over 6 a T} e^{a T -\Delta /T} \;.
\ee

\appendix
\section*{appendix B}\label{app-B}
 
 In this Appendix we obtain explicit expressions for the canonical energy and 
heat  capacity in the independent-particle approximation in terms of 
logarithmic  derivatives of the grand-canonical partition function. 

 We introduce the following notation for the logarithmic derivatives of the
grand-canonical  partition function in the independent-particle model
\begin{equation}
{\cal Z}_{a, b} \equiv {\displaystyle{\partial^{a+b} \ln Z_{GC}
\over \partial \alpha^a \partial \beta^b}}\;.
\end{equation}
Relation (\ref{particle-number}) determines $\alpha =\alpha(\beta,N)$.
Considering  $\alpha$ a function of $\beta$ and $N$ in the relation
(\ref{particle-number}),  we find from  $d N / d\beta = 0$
\begin{equation}
{d \alpha \over d \beta}=-{{\cal Z}_{1,1} \over {\cal Z}_{2,0}}
\end{equation}
We can now use Eqs.~(\ref{partition-N}) and (\ref{E-beta}) to obtain
\begin{equation}\label{energy1}
E_{sp}=-{\cal Z}_{0,1}+{{\cal Z}_{2,1} \over 2 {\cal Z}_{2,0}}-{{\cal
Z}_{3,0}  {\cal Z}_{1,1} \over 2 {\cal Z}_{2,0}^2}
\;.
\end{equation}

Similarly, the canonical heat capacity in the independent-particle model is
given  by
\begin{eqnarray}\label{C-sp}
C_{sp}=-\beta^2\left(-{\cal Z}_{0,2}+{2{\cal Z}_{1,1}^2+{\cal Z}_{2,2} \over
2{\cal  Z}_{2,0}}  - {{\cal Z}_{2,1}^2 +
2{\cal Z}_{3,1}{\cal Z}_{1,1}+{\cal Z}_{3,0}{\cal Z}_{1,2} \over 2{\cal
Z}_{2,0}^2} + {4{\cal Z}_{3,0}{\cal Z}_{2,1}
{\cal Z}_{1,1}+{\cal Z}_{4,0}{\cal Z}_{1,1}^2 \over 2{\cal Z}_{2,0}^3}
-{{\cal  Z}_{3,0}^2 {\cal Z}_{1,1}^2 \over
{\cal Z}_{2,0}^4}\right) \;.
\end{eqnarray}

Eqs.~(\ref{energy1}) and (\ref{C-sp}) are used to calculate the canonical 
thermal  energies and heat capacities in
the  independent-particle model in both the full and truncated spaces.

\appendix
\section*{appendix C}\label{app-C}

 In this Appendix we derive explicit expressions for the
partial  derivatives ${\cal Z}_{a,b}$ of the grand-canonical partition
function (see Appendix B) in terms of the single-particle spectrum and 
scattering  phase-shifts. 

Using Eq.~(\ref{GC-partition}), we find
\begin{equation}\label{Z-ab}
{\displaystyle
{\cal Z}_{a,b}=\sum_{lj} (2j+1) \left[ \sum_{n} (-\epsilon_{nlj})^b
g_{a+b}(\alpha-\beta \epsilon_{nlj})+{1\over \pi}\int_0^\infty d\epsilon
{d\delta_{lj}\over d\epsilon}
(-\epsilon)^b g_{a+b}(\alpha-\beta \epsilon) \right] \;,
}\end{equation}
where we have defined
\begin{equation}
g(x)=\ln (1+e^{x}) \;;\;\;\; g_a \equiv {d^a g \over dx^a}\;.
\end{equation}
Note that $g_1(x)= 1/(1+e^x) =f$, where $f$ is the Fermi-Dirac occupation
number  for $x=\alpha -\beta\epsilon$. We also have
\begin{eqnarray}\label{g-derivatives}
\begin{array}{ll}
g_2=f_1=f(1-f)\\
g_3=f_2=f(1-f)(1-2f)\\
g_4=f_3=f(1-f)(1-6f+6f^2)\\
g_5=f_4=f(1-f)(1-14f+36f^2-24f^3)
\end{array}
\end{eqnarray}

To avoid the numerical derivative of the phase-shift, we can integrate
(\ref{Z-ab})  by parts to obtain
\begin{equation}\label{Z-ab1}
\begin{array}{ll}
{\displaystyle
{\cal Z}_{a,b}=}&{\displaystyle (-1)^b \sum_{lj} (2j+1) \left\{ \sum_{n}
\epsilon_{nlj}^b
g_{a+b}(\alpha-\beta \epsilon_{nlj})
-{b\over \pi}\int_0^\infty d\epsilon \;
\delta_{lj}(\epsilon) \epsilon^{b-1} g_{a+b}(\alpha-\beta \epsilon)\right.
}\\
&{\displaystyle \left.  \;\;\;\;\;\;\;\;\;\;\;\;\;\;\;\;\;\;\;\;\;\;\;\;\;\;\;\;\;\;\;\;\; + {\beta \over \pi}\int_0^\infty d\epsilon \;
\delta_{lj}(\epsilon) \epsilon^b g_{a+b+1}(\alpha-\beta \epsilon)\right\} }
\end{array}
\end{equation}
for $b>0$ and
\begin{equation}
{\displaystyle {\cal Z}_{a,0}=\sum_{lj} (2j+1) \left\{ \sum_{n}
\left[g_a(\alpha-\beta \epsilon_{nlj})-
g_a(\alpha)\right]
+{\beta \over \pi}\int_0^\infty d\epsilon \;
\delta_{lj}(\epsilon) g_{a+1}(\alpha-\beta \epsilon)\right\}}
\end{equation}
for $b=0$.

We note that Eqs. (\ref{particle-number1}) and (\ref{variance1}) can be
rewritten  as
\begin{equation}
N={\cal Z}_{0,1} \;;\;\;\;\;\;\langle (\Delta N)^2\rangle = {\cal Z}_{0,2}
\;,
\end{equation}
where ${\cal Z}_{0,1}$ and ${\cal Z}_{0,2}$ are given as special cases of
(\ref{Z-ab1}).

\end{document}